\documentclass[12pt,a4paper,final]{article}
\author{María Cecilia Tomasini}

\usepackage[top=2.5cm, bottom=3cm, left=2.5cm, right=2.5cm]{geometry}
   
\usepackage[utf8]{inputenc}
\usepackage{amsmath}
\usepackage{amsfonts}

\usepackage{amssymb}
\usepackage{graphicx}
\usepackage{float}

\usepackage{subfig}
\usepackage{graphicx}

\usepackage{caption}
\usepackage{color}

\def\ai{\'{\i}}
\def\lp{\left(}
\def\rp{\right)}
\def\lbr{\left[}
\def\rbr{\right]}
\def\rk{\right\}}
\def\lk{\left\{}

\def\be{\begin{equation}}
\def\ee{\end{equation}}
\def\ba{\begin{array}}
\def\ea{\end{array}}
\def\bea{\begin{eqnarray}}
\def\eea{\end{eqnarray}}

\def\p{\partial}

\def\t{\theta}

\def\lb{\label}
\def\d{\delta}

\def\r{\rho}
\def\m{\mu}
\def\n{\nu}
\def\a{\alpha}
\def\b{\beta}
\def\g{\gamma}
\def\G{\Gamma}
\def\s{\sigma}

\def\om{\omega}
\def\l{\lambda}
\def\P{\Phi}

\def\s{\sigma}

\def\e{\epsilon}

\def\k{\kappa}
\def\z{\zeta}

\def\be{\begin{equation}}
\def\ee{\end{equation}}
\def\ba{\begin{array}}
\def\ea{\end{array}}
\def\bea{\begin{eqnarray}}
\def\eea{\end{eqnarray}}

\begin{document}

\title{\vspace{-1.5cm} \bf Arbitrarily coupled massive scalar field in conical thin-shell spacetimes}
\author{C. Tomasini\footnote{e-mail: ctomasini@df.uba.ar}, E. Rub\ai n de Celis\footnote{e-mail: erdec@df.uba.ar}, C. Simeone\footnote{e-mail: csimeone@df.uba.ar}\\
{\footnotesize Departamento de F\ai sica, Facultad de Ciencias Exactas y Naturales, Universidad de}\\
{\footnotesize Buenos Aires and IFIBA, CONICET, Ciudad Universitaria, Buenos Aires 1428, Argentina.}}
\date{\small \today}

\maketitle
\vspace{-1.0cm} 
\begin{abstract}

We study the massive scalar field of a charged particle held at rest in conical thin-shell spacetimes with one or two asymptotic regions.
Resonant and stable scalar Green's functions are characterized in terms of the coupling of the field to the trace of the extrinsic curvature jump at the shell. Stable coupling values, within the safety domain of the configuration parameters, are used to analyze the self-force of static point charges.

\end{abstract}

\vspace{0.6cm} 

\noindent 
{\small 
PACS number(s): 4.20.-q, 04.20.Gz, 04.40.-b, 41.20.Cv
}

\section{Introduction}
\lb{s1}

Despite their mathematical simplicity, conical spacetimes are of physical interest because they are associated to astrophysical objects which may have appeared in phase transitions in the early Universe and could have played a role in processes of cosmological relevance. 
Spontaneous symmetry breaking in a system including a complex scalar field coupled to a gauge field would lead to the formation of practically one-dimensional topological defects called gauge cosmic strings or local cosmic strings \cite{book}. 
In their most simple configuration, these strings are straight, and they can be described by the approximate energy-momentum tensor 
${\tilde T}_\mu^{\, \nu}= \delta (x)\delta (y) \mbox{diag}(\mathcal{E},0,0,\mathcal{E})$, where $\mathcal{E}$ is the linear mass density \cite{vilenkin3}. 
This density is determined by the energy scale at the moment of the symmetry breaking process. 
From the point of view of the gravitational effects, the equality of the only two non vanishing components of the energy-momentum tensor has the consequence of a metric with $g_{00}=1$, i.e. a null Newtonian force; thus gauge strings could not be detected by their gravitational attraction on rest or non relativistic particles\footnote{The existence of small structures, as wiggles and kinks along the string, would modify this aspect, leading to $g_{00}\neq 1$; see for instance Ref. \cite{vv}.}. 
However, though locally flat, the geometry induced by a straight gauge cosmic string is conical, that is, it presents a deficit angle around its symmetry axis: the line element reads $ds^2=-dt^2+dr^2+(1-8G\mathcal{E})r^2d\theta^2+dz^2$ (see \cite{vilenkin3} and also \cite{hiscock}). 
As a result of this deficit angle, a gauge string deflects light and relativistic massive particles. 
Then such string would act as a gravitational lens which generates double images, and if moving at relativistic speed through a dust cloud or a gas it would generate a matter wake \cite{vilenkin2,turok,sato}. 
The interest on cosmic strings has been mainly driven by the prediction of these observable effects. 
In particular, they have been considered as candidates for the generation of density fluctuations leading to structure formation; while in the last years they have been discarded as the main source for such cosmological process, in the present framework they are considered as possible secondary seeds for structure formation \cite{danos}.

While the absence of effects on rest particles could seem to limit the relevance of conical spacetimes, this aspect radically changes when tests involving charged particles are considered. It is well known that, in a curved background, the electric field of a rest point charge is not spherically symmetric and this leads to the existence of an electrostatic self-force on the charge; 
this is so because though one can always choose a freely falling frame were the Maxwell equations for flat space hold locally, they cannot globally admit the Coulomb solution. 
In particular, then, while a conical spacetime is locally flat, the existence of a deficit angle globally rules out the symmetric electric field solution for a point charge; hence an electrostatic self-force exists and a rest charge would develop an acceleration in such a background \cite{linet}. Besides, and this point is of central importance within our analysis, because the electric field  probes the global aspects of a geometry it also allows to discern between two conical spaces which are identical at the position of the point charge but differ globally. For instance, as shown in Refs. \cite{eoc1,emilio}, one could use the self-force on a charge to identify different possible interior sources behind a shell supporting a conical geometry, or could even detect the existence of a thin-shell wormhole \cite{visser} connecting two conical spacetimes (see for instance Refs. \cite{eiroa,forghani}), without the necessity of traveling across the wormhole throat. Note that wormholes of this class are supported by thin matter layers located at the throat; hence the exotic\footnote{At least within Einstein's theory of gravity; other theories may admit wormhole solutions supported by normal matter; see for instance Refs. \cite{gb1,gb2,gb3}.} matter required for the existence of such geometries is confined at a finite radius and no direct evidence of it is available at the position of the observer using the charge as a test for the spacetime properties.
Since the local curvature vanishes in these conical spacetimes, the self-force has its origin in the boundary conditions on the thin shell and infinity.
The program of computing the electric self-force as a probe of the global properties of spacetime was continued in spherical thin-shell wormholes as well as in thin-shell closed universes (see for instance Refs. \cite{eoc2} and \cite{davidson}). 

The analysis of Refs. \cite{eoc1,emilio} was recently extended \cite{cec17} to address the problem of the self-force on a scalar point charge within the theory of a massless scalar field. 
Scalar fields are considered of interest in modern cosmology, both within the study of the early Universe and, after the first proposals relating the apparently necessary $\Lambda$ term to a scalar field (see for instance Ref. \cite{sasta}), also to understand the present day observations showing an accelerated expansion. 
Scalar fields with non-minimal coupling $\xi$ to gravity have been included in inflationary models
motivated in the early works \cite{Spokoiny,Salopek,Fakir} on the observation
that the problem of an exceedingly small quartic self-coupling parameter, dictated by the amplitude of primordial scalar (density) perturbations \cite{Hawking}, can be solved by using a non-minimally coupled inflaton with a large value of $\xi$.
From a different point of view, a fundamental aspect of the motivation for the study of non-minimally coupled scalar field
is determining the different way in which it behaves in curved geometries, for example, with thin-shell matter sources; while in Maxwell theory the field of a charge is continuous across a non charged matter layer inducing a discontinuity of the extrinsic curvature, for a scalar charge the presence of an infinitely thin matter distribution generates what can be seen as new surface sources for the corresponding coupled scalar field. Thus one cannot, in principle, expect the scalar self-force to behave as the electrostatic self-force.   
The scalar self-force on a static particle is a problem recently addressed in a large class of spherically symmetric spacetimes \cite{bezerra1,popov,popov1,popov2}.
Much of this work was done in wormhole spacetimes, where some authors \cite{taylor1,taylor2} noted that the self-force diverges for an infinite set of values of the curvature coupling $\xi$.
Bezerra and Khusnutdinov \cite{bezerra} analyzed the anomalous behavior through an analogous problem in scattering theory by defining an effective potential in a non-relativistic quantum mechanical problem and identifying the poles of the self-interaction with the bound states for the wave function. 
The study of Ref. \cite{cec17} in cylindrical spacetimes was performed in terms of the coupling of a massless scalar field, and critical values of this constant for which the scalar field is unstable in the background configuration were identified.
In addition, the aforementioned analogy proposed by Bezerra and Khusnutdinov was particularly useful in the latter work to understand the role of the coupling $\xi = 1/4$, which cancels out the delta-like effective potential at the shell in cylindrically symmetric spacetimes, and for which the self-force changes its sign at the vicinities of the shell.

In the present work we generalize our previous results to the case of a massive scalar field. 
Such an extension is interesting for both formal and physical aspects. 
In particular, we well find that, for all other parameters fixed, a greater mass for the scalar field increases the domain of stability in terms of the coupling constant (see Section 3.3). 
From a physical point of view, quantitatively different results are to be expected. 
This can already be inferred from the behavior of the field of a point charge in the simplest situation of a flat spacetime (consider the field equation in Section 2 for the case $R=0$, $\sqrt{-g}=1$): while for a massless field the solution is of the form $1/r$, for a field of mass $m$ the solution takes the form $e^{-mr}/r$. 
Thus, in comparison with the massless case, a more rapidly decaying field is expected, and consequently a weaker self-force. 
In what follows, the problem is studied analytically and numerically in terms of the coupling constant for different configurations of the background spacetime. 
The general formulation of the problem, as well as the treatment of the anomalous behavior pointed above, now for the massive case, are presented in Section 3 in the two types of geometries considered, that is, those associated to cylindrical shells joining an inner to an outer region, and wormholes of the thin-shell class.
The self-interaction on scalar point charges is developed in Section \ref{s4}, and the self-force is studied in Sections \ref{s4.1} and \ref{s4.2} for each type of geometry. Throughout the article the geometrized unit system is used where $c=G=1$.

\section{Approach: massive scalar field coupled to curvature and scalar particle}
\lb{s2}

We consider the action $S = S_{\P}+S_{m_0}+S_{q}$ in a fixed background, where 
\be
S_{\P} = -\frac{1}{ 8 \pi} \int \,d^{4}x \sqrt{-g} \lp g^{\a \b} \p_{\a}\P \,\p_{\b}\P + (m^2 + \xi R)\, \P^{2} \rp \,
\ee
is the action of a massive scalar field with arbitrary coupling $\xi$ to the Ricci curvature scalar $R$, 
\be
S_{m_0} = - m_0 \int_{\gamma} d\tau \,
\ee
is the action of a particle of bare mass\footnote{The bare mass $m_0$ must not be confused with a particular value of the field mass $m$: for example, the latter could be zero if a massless scalar field is considered, while the point particle mass $m_0$ is always non null.} $m_{0}$ over its world line $ \gamma$
and  
\begin{eqnarray}
S_{q} &=& \int_{\gamma} d\tau \, q\,\Phi(y(\tau)) = q \,\int d^{4}x \sqrt{-g}\, \int_{\g} d\tau \,\delta_{4}(x,y(\tau)) \, \Phi(x)
\end{eqnarray}
is the interaction between the field and the particle's charge $q$.
The field equation for a massive scalar field coupled to gravity is derived requiring the action to be stationary under variations $\delta\Phi(x)$. With a scalar charge $q$ at rest the inhomogeneous equation is:
\be \lb{scalar_eq}
\lp \square - \xi R - m^2\rp \P(x) = - 4 \pi q \frac{\d^3 (\bf x - \bf x')}{\dot{y}^{0} \,\sqrt{-g}} \,,
\ee
where $ \delta^{3}(\bf x - \bf x')$ is the spatial Dirac distribution, $\bf x'$ the particle's position, $\square$ the d'Alambertian operator of the metric $g_{\alpha \beta}$, 
and  $\dot{y}^{0}  = dt/d\tau$.

We look for solutions to Eq. (\ref{scalar_eq}) in conical backgrounds with a thin shell. The line element in cylindrical coordinates is
\be \lb{metric}
ds^2 = - dt^2 + dr^2 + \rho^2(r) d\theta^2 + dz^2\, ,
\ee
with $0 < \theta \leqslant 2\pi$, $-\infty<z<+\infty$, and $\rho(r)$ the profile function which determines the type of topology and deficit angles of the conical geometries at each side of the shell. 
Two types will be considered;
\be
\rho(r) = \left\{ \ba{ll}
          \, r \, \om_i \,, & \mbox{in $\mathcal{M}_{i}=\{0<r \leq r_i\}$ }\ \\
\lp r - r_i + r_e \rp \om_e \,, & \mbox{in $\mathcal{M}_{e}= \{ r_i \leq r < \infty \}$}
         \ea \right. 
\ee
for Type I spacetimes, and 
\be
\rho(r) = \left\{ \ba{ll}
        \lp r_- - r \rp \om_- \,, & \mbox{in $\mathcal{M}_{-} = \{ - \infty < r \leq 0\}$ }\ \\
        \lp r + r_+ \rp \om_+ \,, & \mbox{in $\mathcal{M}_{+}= \{0 \leq r < + \infty \}$}
         \ea \right. 
\ee
for Type II.  
In a Type I spacetime the thin shell at $r=r_i$ separates an interior region $\mathcal{M}_{i}$ with an angle deficit $2\pi (1-\om_i)$ from an exterior $\mathcal{M}_{e}$ of angle deficit $2\pi (1-\om_e)$. The first fundamental form over the thin-shell hypersurface relates the internal and external radii by: $\r(r_i) = r_i  \om_i = r_e  \om_e$, \cite{eec,ec}.
Type II geometries correspond to wormhole spacetimes with two asymptotic regions. $\mathcal{M}_{-}$ and $\mathcal{M}_{+}$, characterized by $\om_-$ and $\om_+$ respectively, are joined over the hypersurface of a thin throat through the condition $\r(0) = r_- \om_ - = r_+ \om_+$. In every case, the angle deficit parameter is $0 < \om \leq 1$.
The Ricci scalar is $R(r)= -2 \, \k \, \d(r-r_s)$, where $\k$ is the trace of the jump on the extrinsic curvature tensor over the thin shell \cite{visser,eec} located at the radial position $r_s$; 
\begin{flalign}
\k = \frac{\om_e - \om_i}{\r(r_s)}
\end{flalign}
with $r_s=r_i$ in Type I spacetimes, and  
\begin{flalign}
\k = \frac{\om_+ + \om_-}{\r(r_s)}
\end{flalign}
with $r_s=0$ in Type II.
\\

Going back to the action of the system, the equations of motion for the scalar particle are derived demanding $S$ to be stationary under variations $\d y^{\a}(\tau)$ of the world line, i.e.
\be 
M(\tau) \frac{D \dot{y}^{\a}}{d \tau} = q \lp g^{\a\b} + \dot{y}^{\a} \dot{y}^{\b} \rp \nabla_{\b} \P(y) \,.
\ee
The otherwise dynamical mass $M(\tau) = m_0 - q \P(y)$ is constant if the particle is held at rest in a static spacetime \cite{ppv}. To hold the charge fixed, the total force exerted by a mechanical strut is 
\be
F_{strut}^{\a} = M(\tau) \G^{\a}_{0 0} \dot{y}^{0} \dot{y}^{0} - q \, g^{\a \b} \p_{\b} \P \,.
\ee
The term with the Christoffel symbol $\Gamma^{\a}_{00}$ is zero in the locally flat conical geometries under consideration. The second term corresponds to minus the scalar force and its evaluation needs regularization at the position of the particle. The latter procedure leads to the self-interaction of the scalar particle which will be developed in Section 4.


\section{Green's function for the massive scalar field in conical spacetimes}
\lb{s3}

The field in Type I and Type II spacetimes will be characterized in terms of the configuration parameters by determining resonant or stable Green's functions of the problem. To solve the field equation (\ref{scalar_eq}) we use the three-dimensional static scalar Green's function defined as $\P = 4\pi q G$, thus for either type
{\small
\be \lb{field_eq}  
\lbr \frac{\p^2}{\p r^2} + \frac{\r'(r)}{\r(r)} \frac{\p}{\p r} + \frac{1}{\r^2(r)} \frac{\p^2}{\p \t^2} + \frac{\p^2}{\p z^2 } - \lp \xi\, R(r) + m^2 \rp \rbr G = - \,\frac{\delta^{3}({\bf x} - {\bf x}')}{\r(r)}\,.
\ee
}
We look for a solution as a Fourier expansion in the form
\be\lb{Green_exp}
G = \frac{1}{\pi^2} \sum_{n=0}^{+\infty} \, \frac{\cos[n(\t-\t')]}{1+\d_{0,n} } \int \limits_{0}^{+\infty}dk\,\cos[k(z-z')]\chi_{n}(k,r)
\ee
which yields the following radial equation
\be \lb{radial_eq}
\lk \frac{\p}{\p r} \lbr \r(r) \frac{\p}{\p r} \rbr - \r(r)\lbr \lp \frac{n}{\r(r)} \rp^2 + \xi R(r) + k^2 + m^2 \rbr \rk \chi_n(k,r) =  - \delta(r - r')
\ee
for functions $\chi_{n}(k,r)$, in each $n$-mode and $k$-eigenvalue of the Fourier expansion. 
Integrating (\ref{radial_eq}) over an infinitesimal radial interval around the position of the shell $r=r_s$, we obtain
\be \lb{jump_rs}
\lbr \frac{\p}{\p r} \, \chi_n(k,r) \rbr^{r={r_s}^+}_{r={r_s}^-} = - 2\xi\, \k\, \chi_n(k,r)\big|_{r=r_s} 
\ee
where the Ricci scalar and the continuity of the functions $\r(r)$ and $\chi_n(k,r)$ at $r=r_s$ were used. Analogously, assuming the continuity of $\chi_n(k,r)$ at the radial position $r=r'$ of the source, we have
\be \lb{jump_r'}
\lbr \frac{\p}{\p r} \, \chi_n(k,r) \rbr^{r={r'}^+}_{r={r'}^-} = - \frac{1}{\r(r')} \,.
\ee
The radial solutions $\chi_{n}(k,r)$ will be obtained from (\ref{radial_eq}), with conditions (\ref{jump_rs}) and (\ref{jump_r'}), plus the requirements at the $z$ line axis, or infinity, over the corresponding regions in Type I and Type II geometries.


\subsection{Scalar Green's function in Type I spacetimes}
\lb{s3.1}

In order to obtain the radial solutions of $G$ in Type I spacetimes we define the $\textit{internal}$ and $\textit{external}$ functions as
\be \lb{radial_I}
\chi_n(k,r)= \lk \ba{ll} 
\ba{ll} 
\chi^i
  \quad &\mbox{in} \; \mathcal{M}_{i}  \\
\chi^e
\quad &\mbox{in} \; \mathcal{M}_{e}
\ea 
\ea \right.
\ee
for each $n$ and $k$. 
The boundary conditions at the axis and infinity are
\be \lb{0_inf}
\lim_{r \rightarrow 0} \chi^{i} \neq \infty \,, \quad
\lim_{r \rightarrow +\infty} \chi^{e} = 0 \,.
\ee
These conditions, plus the continuity of the field and its derivative discontinuity (\ref{jump_rs}) at the shell and the conditions for the inhomogeneity (\ref{jump_r'}) at the source's position, determine the radial functions. Considering a source placed in $\mathcal{M}_{i}$, the solutions are
\begin{flalign} 
\chi^{i} = \, \chi^{\om_i} + \frac{1}{\om_i} I_{\l}(\m \r(r)/\om_i )  \, A_n(k) \;\;\; \mbox{with } \l = n/\om_i \,, \lb{radial_i}\\
\chi^{e}  = \, \frac{1}{\om_i} K_{\n}(\m \r(r)/\om_e) \,C_n(k) \;\;\;  \mbox{with } \n = n/\om_e \,,  
\end{flalign}
where $\mu = \sqrt{k^2 + m^2}$, and the terms $\chi^{\om_i}$ are given by
\begin{flalign} \lb{rad_cs_i}
\chi^{\om_i} =  \, \frac{1}{\om_i} I_{\l}(\m \r(r_<)/\om_i) K_{\l}(\m \r(r_>)/\om_i)  \,,  \; \ba{ll} 
r_< = \mbox{min}\{ r, r' \} \\
r_> = \mbox{max}\{ r, r' \}  
 \ea \,.
\end{flalign}
The functions $\chi^{\om_i}$ satisfy the jump (\ref{jump_r'}) and account for the source inhomogeneity of $\chi^{i}$ in the internal region, while the second terms in (\ref{radial_i}) are smooth at the position $r=r'$.
Finally, the coefficients $A_n(k)$ and $C_n(k)$ determined from the boundary conditions at the shell are, explicitly:
\begin{flalign} \lb{A_n}
A_{n}(k) =  - I_{\l}(\mu r')  \times \frac{K_{\l}(\mu r_i) K'_{\n}(\mu r_e) - K_{\n}(\mu r_e) K'_{\l}(\mu r_i) + 2 \k \xi K_{\n}(\mu r_e) K_{\l}(\mu r_i)}{I_{\l}(\mu r_i) K'_{\n}(\mu r_e) - K_{\n}(\mu r_e) I'_{\l}(\mu r_i) + 2 \k \xi K_{\n}(\mu r_e) I_{\l}(\mu r_i)}
\end{flalign}
\begin{flalign} \lb{C_n}
C_n(k) = 
\frac{A_{n}(k) I_{\l}(\mu r_i) + I_{\l}(\mu r')K_{\l}(\mu r_i)}{K_{\n}(\mu r_e)} 
\end{flalign}
with the prime over Bessel functions implying a derivative with respect to the corresponding radial coordinate.

If the source is placed in region $\mathcal{M}_{e}$, the radial solutions are
\begin{flalign}
\chi^{i} =  \frac{1}{\om_e} I_{\l}(\m \r(r)/\om_i) \, D_n(k) \;\;\; \mbox{with }\l = n/\om_i \,,  \\
\chi^{e}  = \, \chi^{\om_e} + \frac{1}{\om_e} K_{\n}(\m \r(r)/\om_e) \, B_n(k) \;\;\; \mbox{with } \n = n/\om_e \,. \lb{radial_e}  
\end{flalign}
The terms $\chi^{\om_e}$ which account for the inhomogeneity in the external region are given by
\begin{flalign} \lb{rad_cs_e}
\chi^{\om_e} =  \, \frac{1}{\om_e} I_{\n}(\m \r(r_<)/\om_e) K_{\n}(\m \r(r_>)/\om_e)  \,,  \; \ba{ll} 
r_< = \mbox{min}\{ r , r' \} \\
r_> = \mbox{max}\{ r, r' \}  
\ea 
\end{flalign}
and the coefficients are
{\small
\begin{flalign} \lb{B_n}
B_{n}(k) =  - K_{\n}((r'-r_i+r_e)\m)   \frac{I_{\l}(\mu r_i) I'_{\n}(\mu r_e) - I_{\n}(\mu r_e) I'_{\l}(\mu r_i) + 2 \k \xi I_{\n}(\mu r_e) I_{\l}(\mu r_i)}{I_{\l}(\mu  r_i) K'_{\n}(\mu  r_e) - K_{\n}(\mu  r_e) I'_{\l}(\mu  r_i) + 2 \k \xi K_{\n}(\mu  r_e) I_{\l}(\mu  r_i)} 
\end{flalign}
}
\begin{flalign} \lb{D_n}
D_n(k) = 
\frac{I_{\n}(\mu  r_e) K_{\n}((r'-r_i+r_e)\m) + B_{n}(k) K_{\n}(\mu  r_e)}{I_{\l}(\mu r_i)} \,.
\end{flalign}

We point out that the terms $\chi^{\om} = \chi^{\om}_{n}(k,r)$ (with $\om=\om_i$ or $\om=\om_e$) are intentionally left apart because the Green's function for a profile $\rho(r) = r \om$, which corresponds to an infinitely thin straight cosmic string manifold without a shell ($0< r < \infty$), is 
\be\lb{exp_cs}
G_{\om} = \frac{1}{\pi^2} \sum_{n=0}^{+\infty}  \, \frac{\cos[n(\t-\t')]}{1+\d_{0,n}} \int \limits_{0}^{+\infty}dk\,\cos[k(z-z')]	\, \chi^{\om}_{n}(k,r)  \,.
\ee
This is the scalar Green's function in a spacetime with a conical-type line singularity which is known in an integral form \cite{guimaraes} as
\begin{flalign} \lb{Green_cs}
G_{\om}  = \frac{1}{4 \pi^2 \om \sqrt{2 r r'}} \int \limits_{u}^{+\infty}  d\z \, \frac { \cos{\lp m \sqrt{2 r r'} \sqrt{\cosh \z - \cosh u } \rp} \;  \sinh(\z/\om) }{\sqrt{\cosh \z - \cosh u} \lp \cosh(\z/\om) - \cos{(\t- \t')} \rp } 
\end{flalign} 
with
\be
\cosh u = \frac{r^2 + r'^2 + (z-z')^2}{2 r r'} \,, \quad u  \geqslant 0 \,.  \nonumber
\ee
We will use this result to write the actual Green's function (\ref{Green_exp}), over the region where the source is placed, as the sum of two parts, i.e.
\be  \lb{G Type I}
G = G_{\om} + G_{\xi} \quad \mbox{if $(x, x') \in \mathcal{M}_i$ or $\mathcal{M}_e$}\,
\ee
where $\om = \om_i$ (or $\om = \om_e$). The second part, $G_{\xi}$, is homogeneous in $\mathcal{M}_i$ (or $\mathcal{M}_e$) and is constructed from the second terms in (\ref{radial_i}) (or (\ref{radial_e})). With this splitting we have $G_{\om}$ accounting for the source inhomogeneity, while $G_{\xi}$ contains the information about the coupling $\xi$ and the specific boundary conditions of the thin-shell spacetime. \\


\subsection{Scalar Green's function in Type II spacetimes}
\lb{s3.2}

For Type II geometries, corresponding to wormholes with two asymptotic regions, we define the $\textit{minus}$ and $\textit{plus}$ radial functions of $G$ as
\be \lb{radial_II}
\chi_n(k,r)= \lk \ba{ll} 
\ba{ll} 
\chi^-
  \quad &\mbox{in} \; \mathcal{M}_{-}  \\
\chi^+
\quad &\mbox{in} \; \mathcal{M}_{+}
\ea 
\ea \right.
\ee
for each $n$ and $k$. 
The boundary conditions are
\be \lb{inf_-+}
\lim_{r \rightarrow -\infty} \chi^{-} = 0 \,, \qquad
\lim_{r \rightarrow +\infty} \chi^{+} = 0 \,
\ee
at the infinities, continuity of the field and its derivative discontinuity (\ref{jump_rs}) at the shell, 
and the inhomogeneity discontinuity (\ref{jump_r'}) at the source's position $r=r'$. Considering a source placed in region $\mathcal{M}_{+}$, the radial solutions are
\begin{flalign}
\chi^{-} = \frac{1}{\om_+} K_{\l}(\m \r(r)/\om_-)  \, E_n(k) \;\;\; \mbox{with } \l = n/\om_- \,, \lb{radial_wh_-}\\
\chi^{+}  =  \, \chi^{\om_+} + \, \frac{1}{\om_+} K_{\n}( \m \r(r)/\om_+) \, W_n(k)\;\;\; \mbox{with } \n = n/\om_+ \,, \lb{radial_wh_+}
\end{flalign}
where the terms which account for the inhomogeneity at $r=r'$ are
\begin{flalign}
\chi^{\om_+} =  \, \frac{1}{\om_+} I_{\n}( \m \r(r_<)/\om_+) K_{\n}(\m \r(r_>)/\om_+)  \,,  \; \ba{ll} 
r_< = \mbox{min}\{ r, r' \} \\
r_> = \mbox{max}\{ r, r' \}  
\ea .
\end{flalign}
The coefficients $E_n(k)$ and $W_n(k)$, determined from the boundary conditions at the shell, are
{\small
\begin{flalign} \lb{W_n}
W_n(k) =  -  K_\n((r' + r_+) \m) 
&\frac{I_{\n}(\mu  r_+) K'_{\l}(\mu r_-) + K_{\l}(\mu  r_-)I'_{\n}(\mu  r_+) + 2 \k \xi I_{\n}(\mu  r_+) K_{\l}(\mu  r_-)}{K_{\l}(\mu  r_-) K'_{\n}(\mu  r_+) + K_{\n}(\mu  r_+) K'_{\l}(\mu  r_-) + 2 \k \xi K_{\n}(\mu  r_+) K_{\l}(\mu  r_-)} 
\end{flalign}
}
\begin{flalign} \lb{E_n}
E_n(k) =  
\frac{I_{\n}(\mu  r_+)K_{\n}((r' + r_+) \m) + W_{n}(k) K_{\n}(\mu  r_+)}{K_{\l}(\mu r_-)} 
\end{flalign}
with the prime over Bessel functions implying a derivative with respect to the corresponding radial coordinate.
In the same way that it was done with Type I solutions, we can separate 
\be \lb{G Type II}
G = G_{\om_+} + G_{\xi} \quad \mbox{if $(x, x') \in \mathcal{M}_+$}\,,
\ee
where $G_{\om_+}$, inhomogeneous at $x'$, is constructed with the radial solutions $\chi^{\om_+}$ and is locally equivalent to the integral solution (\ref{Green_cs}) with $\om=\om_{+}$, while $G_{\xi}$ is constructed with the second terms of (\ref{radial_wh_+}), is regular at $x'$ and contains the information about the coupling and boundary conditions.


\subsection{Stability regions and resonant configurations}
\lb{s3.3}

The Fourier series solutions (\ref{Green_exp}) for the scalar Green's function in Type I and Type II spacetimes are given in terms of coefficients (\ref{A_n})-(\ref{C_n}), (\ref{B_n})-(\ref{D_n}) or (\ref{W_n})-(\ref{E_n}), obtained in the previous subsections. These coefficients may diverge at some eigenvalue $k=k_p$ for specific values of the coupling constant, as can be seen from their denominators in expressions (\ref{A_n}), (\ref{B_n}) and (\ref{W_n}). The coupling $\xi_p^{(n)}$ for which the $n^{th}$ coefficient diverges at $k=k_p$ is
\be
\xi_p^{(n)} = \frac{ K_{\n}(\mu  r_e) I'_{\l}(\mu  r_i) - I_{\l}(\mu  r_i) K'_{\n}(\mu  r_e)}{2 \k K_{\n}(\mu  r_e) I_{\l}(\mu  r_i)} \Big|_{k=k_p} \quad \mbox{in Type I, } 
\ee
and
\be
\xi_p^{(n)} = - \frac{K_{\l}(\mu  r_-) K'_{\n}(\mu  r_+) + K_{\n}(\mu  r_+) K'_{\l}(\mu  r_-) }{2 \k K_{\n}(\mu  r_+) K_{\l}(\mu  r_-)}\Big|_{k=k_p}  \quad \mbox{in Type II,} 
\ee
where the dependance on $k$ appears in $\mu = \sqrt{k^2 + m^2}$. 
This means that a Fourier integral in $dk$ for the $n^{th}$ mode of the Green's function presents a pole at $k=k_p>0$ if $\xi = \xi_p^{(n)}$. Nevertheless, the integral can be performed by splitting it at $k_p \mp \e$, with $\e \to 0$, canceling out the contributions of the lateral limits in the positive real line integral or, equivalently, using contour integrals for the complex-valued coefficient function to obtain the positive real half-line integral.
On the other hand, a mode is divergent if the denominator of the coefficient in its integrand is null for $k = 0$. 
Then, if a specific value of the coupling makes the $n^{th}$ coefficient divergent at $k=0$, the $n^{th}$ mode in (\ref{Green_exp}) is resonant and the massive coupled scalar field in the given conical thin-shell background has an unstable configuration.
The critical coupling $\xi_c^{(n)} = \xi_p^{(n)} \big|_{k=0}$ that generates a resonant $n^{th}$ mode is:
\begin{flalign} 
\xi_c^{(n)} 
=  \frac{ K_{\n}(m  r_e) I'_{\l}(m r_i) - I_{\l}(m r_i) K'_{\n}(m r_e)}{2 \k K_{\n}(m r_e) I_{\l}(m r_i)}  \quad \mbox{in Type I,}
\end{flalign}
and
\begin{flalign} 
\xi_c^{(n)} 
= - \frac{K_{\l}(m  r_-) K'_{\n}(m r_+) + K_{\n}(m r_+) K'_{\l}(m r_-) }{2 \k K_{\n}(m r_+) K_{\l}(m r_-)}  \quad \mbox{in Type II.} 
\end{flalign}
To study the problem of a coupled scalar field and, in particular, the self-interaction of a scalar particle in a given background, it is convenient to work with solutions which do not present poles in $k$ for the Fourier integral modes. 
Moreover, to analyze the dependence of the self-force in terms of the coupling constant we would like to go over values of $\xi$ in an interval where we do not come upon resonant configurations. 
To find ranges of the coupling constant where no resonant configuration or poles are found, we will look for safety domains in the parameter space of the configuration.

\begin{figure} [h!] 
\centering
\captionsetup[subfigure]{margin={0.8cm,0.5cm},justification=centering}
\subfloat[Negative $\k$ thin-shell configuration is stable for $\xi$$>$$\xi_{c}^{(0)}$.]
{\label{f1a}\includegraphics[width=0.49\textwidth]{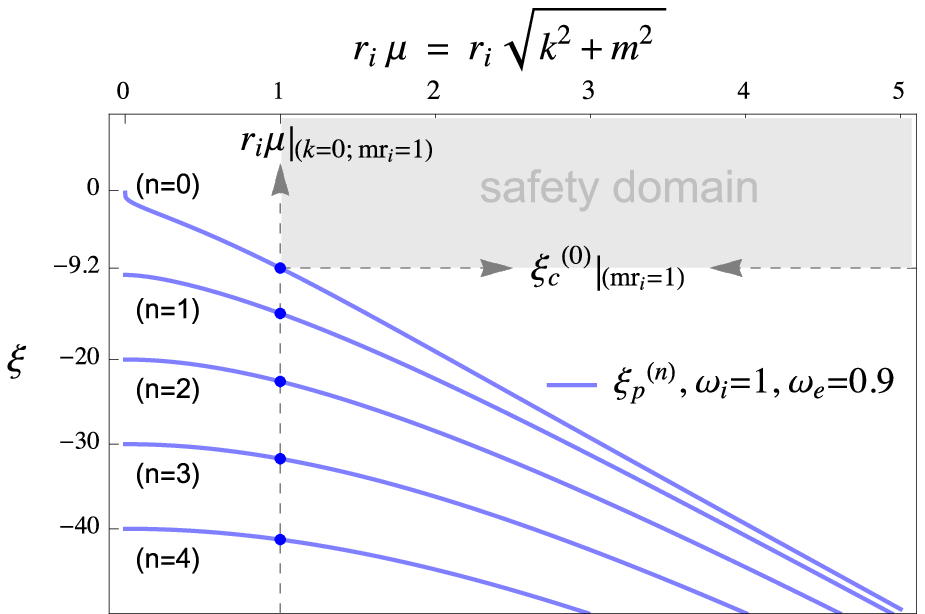}}
 \hspace{0.01cm}
\subfloat[Positive $\k$ thin-shell configuration is stable for $\xi$$<$$\xi_{c}^{(0)}$.]
  {\label{f1b}
    \includegraphics[width=0.49\textwidth]{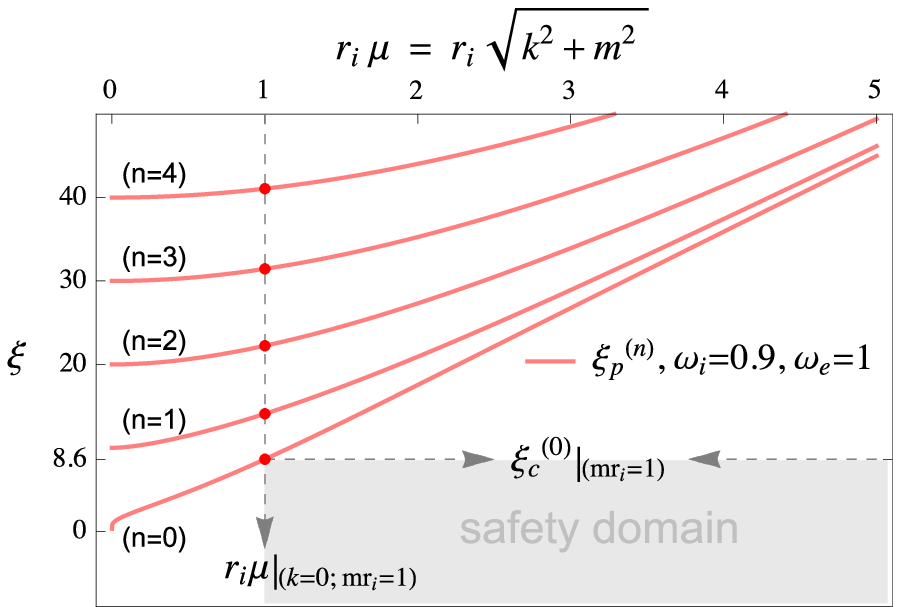}}
    \caption{{Safety domains, in grey, for the massive coupled scalar field Green's function in Type I spacetimes with $m r_i = 1$. The curves represent the function $\xi_{p}^{(n)}$ for different $n$ modes.}
    } 
      \label{f1} 
    \end{figure}

Figure \ref{f1} shows the plot of $\xi_{p}^{(n)}$ against $r_{i}\m$ for Type I spacetimes; in the example of Figure \ref{f1a} the trace of the extrinsic curvature jump $\k$ is negative, and in Figure \ref{f1b} $\k$ is positive.
Each point in a curve $\xi_{p}^{(n)}$ represents a divergence of the $n^{th}$ coefficient for the corresponding values of $\xi$ and $r_i \mu$ at eigenvalues $k$. To analyze a particular configuration one must fix the product $m r_i$, and look for the critical couplings at the intersection of the curves with the vertical line ${r_i \mu}|_{k=0} = m r_i$. To the right side of this line the curves represent values of the coupling for which poles appear in the corresponding $n$ coefficient for some eigenvalue $k>0$.
In Figure \ref{f1a} a safety domain is painted in grey for a Type I geometry with $\om_i=1$, $\om_e=0.9$ ($\k<0$) and $m r_i=1$. For this configuration, $\xi_{c}^{(0)}|_{m r_i =1} \simeq -9.2$ is the greatest critical coupling, thus $\xi > \xi_{c}^{(0)}|_{m r_i =1}$ is a stable range where no resonant configurations are encountered. There is no intersection of the curves $\xi_{p}^{(n)}$ with the grey region; this ensures that while increasing the eigenvalue $k$, with fixed $m r_i=1$, no poles appear for any coefficient $n$ in the safety domain.
Figure \ref{f1b} shows, in grey, a safety domain for a Type I geometry with $\om_i=0.9$ and $\om_e=1$ ($\k>0$), where we chose again $m r_i=1$. In this case $\xi_{c}^{(0)}|_{m r_i =1} \simeq 8.6$ is the smallest critical coupling and $\xi < \xi_{c}^{(0)}|_{m r_i =1}$ is the stable region for configurations with positive $\k$ thin-shells.

Analogously, for a Type II spacetime with given parameters $\om_{\mp}$ and throat radius $r_0$, we can plot $\xi_{p}^{(n)} $ against $r_{0}\mu$. Two cases are presented in Figure \ref{f2}, a symmetric thin-shell Minkowski wormhole ($\om_{\mp} = 1$) and a conical one ($\om_{\mp}=0.9$), where $r_0 = (r_-+r_+)/2 = r_{\mp}$. The safety domain in grey is valid for both configurations with fixed $m r_0=1$. Couplings $ \xi <  \xi_{c}^{(0)}|_{m r_0 =1} \simeq 0.7$ are stable, and no poles are found in the grey region where there is no intersection with the curves $\xi_{p}^{(n)}$. 
\begin{figure}  [H]
 \centering
   \includegraphics[width=0.50\textwidth]{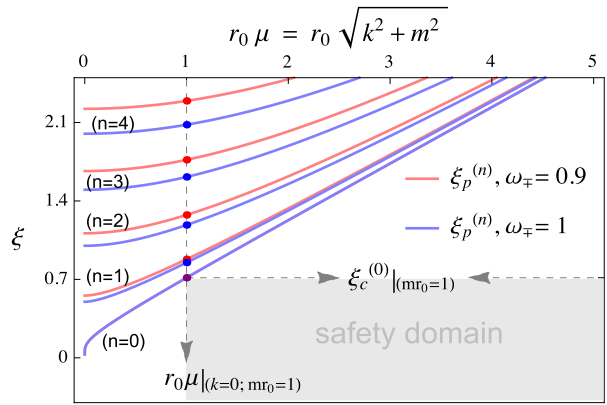}
   \caption{Safety domain, in grey, for the massive coupled scalar field Green's function in symmetric thin-shell wormhole backgrounds with $m r_0 = 1$. If $\xi$$<$$\,\xi_{c}^{(0)}$ the configuration is stable.}
      \lb{f2}
\end{figure}

As we have seen in the previous examples, $\,\xi_{c}^{(0)}$ is always the largest (smallest) critical coupling if $\k$$<$$0$ ($\k$$>$$0$) and therefore determines the boundary of stability of the solutions.  
Moreover, there is no upper (lower) bound, so by restricting $\xi$$>$$\,\xi_{c}^{(0)}$ ($\xi$$<$$\,\xi_{c}^{(0)}$) the configuration with negative (positive) $\k$ is stable.
Either for Type I or Type II spacetimes the stable range for $\xi$ is larger as long as the product $m r_i$ or $m r_0$ is increased. The latter can be observed from Figures \ref{f1} and \ref{f2} or, in general, by looking the following expansions
\be
\xi_p^{(n)} \simeq \frac{\om_i}{\om_e - \om_i} \, \mu r_i  + \mathcal{O}\lp\frac{1}{\mu r_i}\rp \approx \; \frac{\m}{\k}
\,,\quad \mbox{$\mu \gg r_i^{-1}$ in Type I,}
\ee
\be
\xi_p^{(n)} \simeq \frac{2 \om_- \om_+}{(\om_+ + \om_-)^2} \, \mu r_0+ \frac{1}{4} + \mathcal{O}\lp\frac{1}{\mu r_0}\rp \approx  \; \frac{\m}{\k}
\,, \quad \mbox{$\mu \gg r_0^{-1}$ in Type II.}
\ee
Keeping the spacetime parameters fixed, a greater mass for the scalar field increases the absolute value $|\xi_p^{(n)}|$. Particularly, for $m \gg r_i^{-1}$ (or $m \gg r_0^{-1}$) we have $\xi_c^{(n)} \sim m/\k$, i.e. a greater mass enlarges the range of stable values for the coupling constant.
On the other hand, the stable range of the coupling decreases if the mass of the scalar field is small compared to the reciprocal of the shell's radius, namely, to zeroth order 
\be
\xi_c^{(n)} 
\approx \frac{n}{\k \, \r(r_s)}
\,, \quad \mbox{ for $m \ll r_{i}^{-1}$ or $m \ll r_{0}^{-1}$
.}
\ee
With $m\to0$, then $\xi_c^{(0)} \to 0$ and the safety domain takes positive or negative values of $\xi$ depending on whether the shell has $\k<0$ or $\k>0$, respectively. In this limit we recover the critical couplings $\xi_c^{(n>0)} = n (\k \, \r(r_s))^{-1}$ for massless scalar fields in conical thin-shell spacetimes \cite{cec17}.


\section{Scalar charge self-interaction}
\lb{s4}

The self-interaction of the particle is evaluated with a regular field obtained by subtracting the singular part from the actual Green's function in a neighborhood of the charge's position $\bf x'$.
The singular scalar Green's function is constructed with the Detweiler-Whiting procedure \cite{DW}; for a static particle in a static locally flat spacetime it is simply
\be \lb{G_DW} 
G_{DW}
 = \frac{1}{4 \pi} \lp \frac{1}{\sqrt{2\s}} - m \rp + \mathcal{O} (\s^{1/2})
\,,
\ee
over the neighborhood of $\bf x'$, 
where $\s=\s( \bf x, x')$ is half the squared geodesic distance between $\bf {x}$ and $\bf {x}'$ as measured in the purely spatial sections of the spacetime \cite{popov}. The regular function is then $G_R = G - G_{DW} $.

As we have seen in expressions (\ref{G Type I}) and (\ref{G Type II}), the Green's function in the region where the particle is placed is $G = G_{\om} + G_{\xi}$, with the corresponding parameter $\omega$. 
The inhomogeneous part $ G_{\om}$ is divergent at the charge's position and can be further split as the sum of a singular and a regular part. To do this we take the local form used by Guimaraes and Linet in \cite{guimaraes}, valid for points $\bf {x}$ close enough to $\bf {x}'$ and over the range $\pi(1 - 2 \om )< \t - \t' <  \pi (2 \om - 1)$ of the angular coordinate, written as
\be \lb{Guimaraes}
G_{\om} \simeq \frac{1}{ (2 \pi)^{3/2} } \lp \frac{m}{d} \rp^{1/2} K_{\frac{1}{2}}(m d) + G^{reg}_{\om} 
\ee
where $d=\sqrt{r^2 + r'^2 - 2rr' \cos{(\t-\t')} + (z-z')^2}$ is the Euclidean distance, and the regular part is given by
\begin{flalign} \lb{Green_cs_reg}
G^{reg}_{\om} =  \frac{1}{\om} \frac{m^{1/2}}{(2 \pi)^{5/2}} \int \limits_{0}^{+\infty} d\z \, K_{\frac{1}{2}}(D(\z)m) \frac{F_{\om}(\z, \theta - \theta')}{\sqrt{D(\z)}}     
\end{flalign} 
with $D(\z)=\sqrt{ r^2 + r'^2 + 2rr' \cosh \z + (z-z')^2}$,  and 
$$
F_{\om}(\z, \psi) = \frac{\sin (\psi - \pi / \om)}{\cosh (\z / \om) - \cos (\psi - \pi / \om)} - \frac{\sin (\psi + \pi / \om)}{\cosh (\z / \om) - \cos (\psi + \pi / \om)} \,.
$$
The first term in (\ref{Guimaraes}) is exactly the singular part of the Green's function for a massive scalar field in a three-dimensional conical geometry. The latter can be seen from its Taylor expansion  
\begin{flalign} \lb{G_S}
G_S \equiv & \; \frac{1}{(2 \pi)^{3/2}} \lp \frac{m}{d} \rp^{1/2} K_{\frac{1}{2}}(m d)  \nonumber\\
= & \; \frac{1}{4 \pi} \lp \frac{1}{d} - m \rp + \mathcal{O} (d^{1})
\end{flalign}
which shows the structure of the singular Green's function in Euclidean space and the mass term corresponding to the massive field as in (\ref{G_DW}). Finally, the regular Green's function in a neighborhood of ${\bf x'}$ is,
\begin{flalign}
G_R = G - G_{DW} = G^{reg}_{\om} +  G_{\xi} \,.
\end{flalign}
The regular field evaluated at ${\bf x'}$ is $\Phi_R\big|_{x'} =  4\pi q G_R\big|_{x'} $, and the self-energy of the particle is $U_{self}=\dfrac{q}{2}\,\Phi_R\big|_{x'} $. The self-force is obtained from the derivative of the regular field. Due to the cylindrical symmetry of spacetime, the procedure yields a radial force $f  = f_{\om} + f_{\xi}$, where
\begin{flalign} 
f_{\om} =  4 \pi \, q^2\, \frac{\p G_{\om}^{reg}}{\p r}   \Big|_{x'}  \quad \mbox{      and      } \quad 
f_{\xi}  =  4 \pi \, q^2\, \frac{\p G_{\xi}}{\p r}  \Big|_{x'}   \,. \lb{force_xi} 
\end{flalign} 
In the following subsections this is applied to evaluate the self-force on a scalar charged particle in Type I and Type II spacetimes.

\subsection{Self-force in Type I geometries}
\lb{s4.1}

\begin{figure} [h!] 
\centering
\captionsetup[subfigure]{margin={0.9cm,0.0cm},justification=centering}
\subfloat[Negative $\k$ thin shell ($\om_i=1,\, \om_e=0.9$)]
{\label{sfTIord}\includegraphics[width=0.44\textwidth]{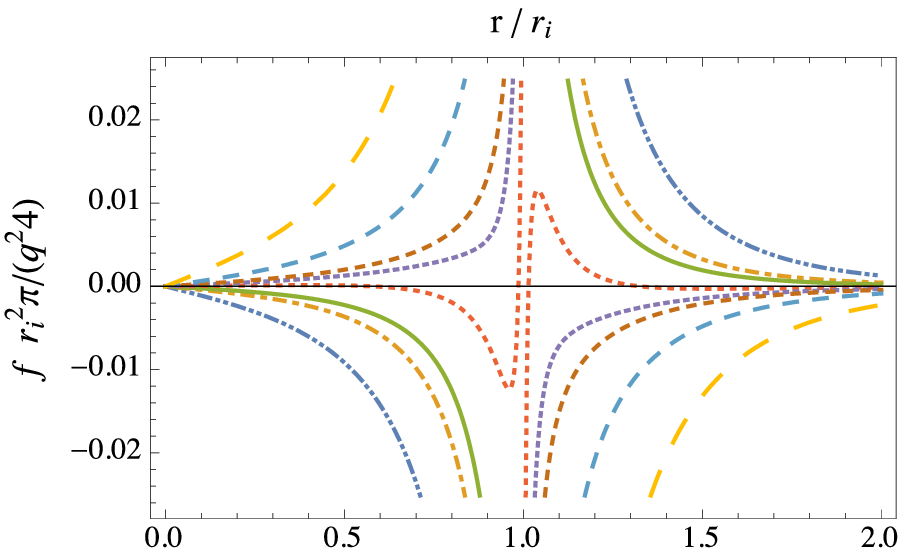}} 
\subfloat[Positive $\k$ thin shell ($\om_i=0.9,\,\om_e=1$)]
  {\label{sfTIexot}
    \includegraphics[width=0.44\textwidth]{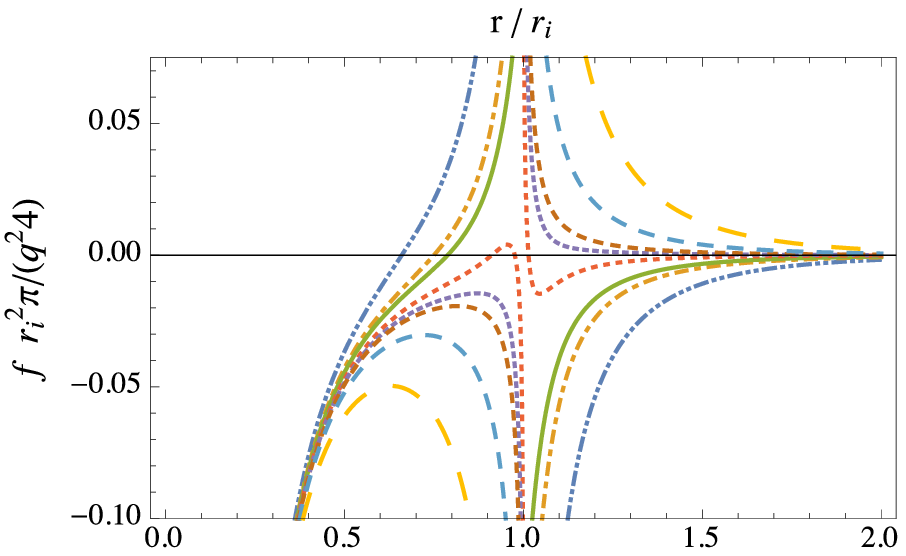}}
{\label{}\includegraphics[width=0.14\textwidth]{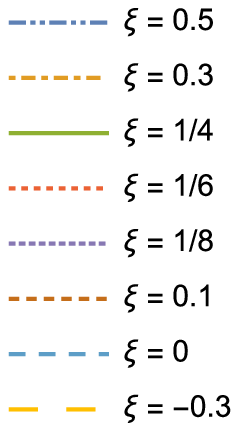}}
    \caption{{Dimensionless self-force $\frac{\pi r_i^2}{4q^2} f$ as a function of $r/r_i$ in a Type I spacetime with $r_i m = 1$ and different coupling constants.
    }
    } 
      \label{sf_T1} 
    \end{figure}

\begin{figure} [h!] \centering
\captionsetup[subfigure]{margin={0.9cm,0.0cm},justification=centering} \subfloat[Minimal coupling] 
{\label{Type_1_masa_xi_cero}\includegraphics[width=0.49\textwidth]{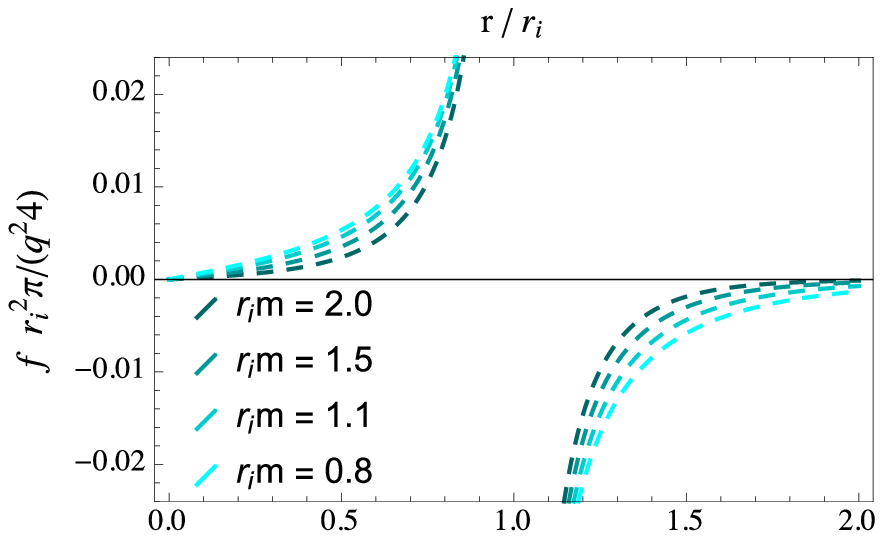}}
\hspace{0.01cm}
\subfloat[$\xi=0.3$] 
{\label{Type_1_masa_xi_03} \includegraphics[width=0.49\textwidth]{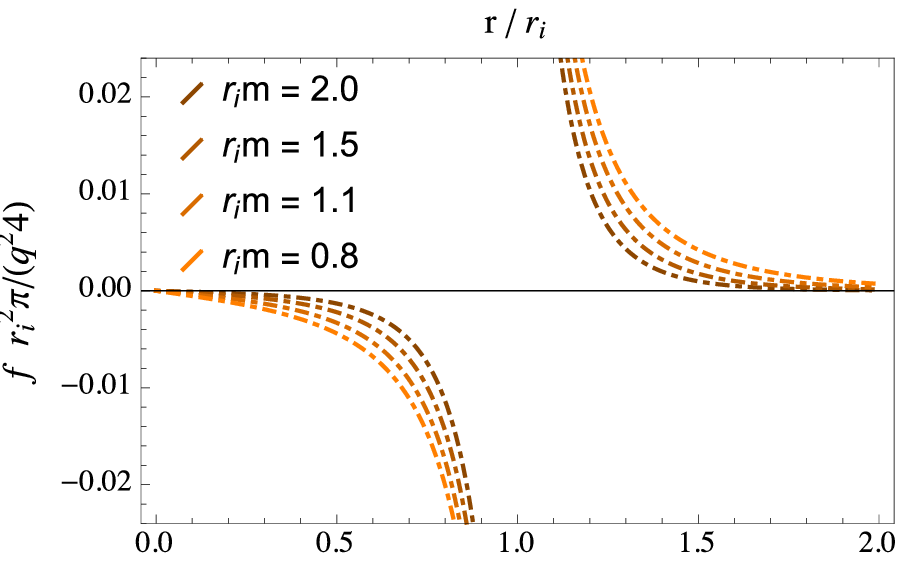}}
\caption{{Dimensionless self-force $\frac{\pi r_i^2}{4q^2} f$ as a function of $r/r_i$ in a Type I spacetime with $ \kappa < 0 $ ($\om_i=1$, $\om_e=0.9$) and different values of $r_i m$.}
    }    \label{f_masa_ord} 
\end{figure}

The self-force $f$ over the scalar particle in a Type I geometry with $ \omega = \omega_{i}$ in the interior region and $\omega = \omega_{e}$ in the outer one is given by
\begin{flalign} 
f_{\om} 
= 
\frac{q^2}{4 \pi} \frac{\sin{(\pi/\om)}}{\lp \r(r')/\om \rp^2}  \,\frac{1}{\om}
\int \limits_{0}^{+\infty} d\z \; \frac{ 1 + \a(\z)  }{ \cosh{\frac{\z}{\om}} -\cos{\frac{\pi}{\om}} } \; \frac{ e^{-\a(\z)} }{\cosh{\frac{\z}{2}}}   \lb{f_w_I} 
\end{flalign}
where $\a(\z) = 2 m \,\frac{\r(r')}{\om}\,\cosh{\frac{\z}{2}}$, and
\begin{flalign} \lb{f_xi_I}
f_{\xi} = \left\{\ba{ll} 
\frac{4q^2}{\pi w_i} 
\sum\limits_{n=0} ^{+\infty} \frac{1}{1+\delta_{n,0}} \int \limits_{0}^{+\infty} dk \, I'_{\l}(r' \sqrt{k^2+m^2}) A_{n}(k) \,, &\mbox{if $r'<r_i$} \\
\frac{4q^2}{\pi w_e}
 \sum\limits_{n=0}^{+\infty} \frac{1}{1+\delta_{n,0}}
 \int \limits_{0}^{+\infty}
 dk \, K'_{\n}\lp (r'-r_i+r_e) \sqrt{k^2+m^2} \rp B_{n}(k)
\,,  &\mbox{if $r'>r_i$}
\ea \right. &
\end{flalign}
The second term in $f = f_{\om} + f_{\xi}$ is obtained from $G_{\xi}$, using the smooth terms of the radial solutions (\ref{radial_i}) and (\ref{radial_e}).
The results are shown in figures \ref{sf_T1} and \ref{f_masa_ord} in which the dimensionless self-force $ \frac{\pi r_{i}^{2}}{4 q^{2}} \, f$ is plotted against the dimensionless position $r/r_{i}$ of the scalar particle for different values of the coupling constant $ \xi $ and spacetime parameters.


Figure \ref{sfTIord} shows the self-force in a Type I geometry with an ordinary matter thin shell of $\kappa < 0$.
The parameters in this configuration are $\omega_{i}=1$, $ \omega_{e} = 0.9$ and $r_i m = 1$.
The coupling constants have been chosen within the stability range according to Figure \ref{f1a}.
In every case, the force becomes divergent as approaching to the shell's position, i.e., $r \rightarrow r_{i}$. 
The pure effect of a shell with $\k<0$ is observed at minimal coupling for which the force is attractive towards the shell.
A negative value of $\xi$ coupled to negative $\k$ increases the attraction to the shell. Conversely, a positive $\xi$ with negative $\k$ contributes repulsively; a weakened attractive force is seen if $\xi<1/6$, and a dominant repulsion from the shell is observed everywhere if $\xi \geqslant 1/4$.
The value $\xi = 1/4$ was identified in a previous work \cite{cec17} as the coupling which cancels out the delta-like potential at the shell in the analogous problem (for the radial solutions) defined in terms of a wave function propagating in an effective potential.
Figure \ref{sfTIexot} refers to a Type I spacetime with an exotic matter thin shell, $ \kappa > 0 $, of parameters $ \omega_{i}=0.9 $, $ \omega_{e}=1$ and $r_i m = 1$. 
The values of the coupling constant
have been chosen within the stability range of Figure \ref{f1b}. 
In the interior region the particle is attracted to the center due to the conical singularity. 
For minimal coupling the charge is repelled at the vicinities of the thin shell with $\k>0$. 
A negative $\xi$ coupled to positive $\k$ increases the repulsion, while positive $\xi$ and positive $\k$ contributes attractively. The behavior for $r \to r_i$ is reversed at $\xi = 1/4$ and, in this case, the force becomes attractive for $\xi \geqslant 1/4$ at the vicinities of the shell.

Finally, Figure \ref{f_masa_ord} shows the dependance of the force on the mass by fixing different values of the product $r_i m$. We observe that the self-force is attenuated with increasing $r_i m$ at both sides of the shell.
In every case the self-force decreases with increasing mass as expected from the exponential decay $e^{-mr}/r$ in the general solution of the field in these conical geometries. 
Although this specific singular term is subtracted as seen from Eq. (\ref{G_S}) in the regularization procedure, traces of the same dependance are found in the regular term $G_{\om}^{reg}$, where $K_{1/2}(m D(r))/\sqrt{r} \sim e^{-mr}/r$. 
Of greater difficulty is interpreting exactly the same factor in the contribution $G_{\xi}$ to the Green's function. We note that in regions where $\om=1$, the derivative of the latter is the only term contributing to the self-force, and the same kind of decay and attenuation with increasing $r_i m$ is observed, as shown in the inner region of Figures \ref{Type_1_masa_xi_cero} and \ref{Type_1_masa_xi_03}.


\subsection{Self-force in Type II geometries}
\lb{s4.2}

The self-force on a scalar particle placed in the region $\mathcal{M}_+$ of a Type II geometry is given by $f = f_{\om_+} + f_{\xi}$ with
\begin{flalign} 
f_{\om_+} 
= \frac{q^2}{4 \pi} \frac{\sin{(\pi/\om_+)}}{(r'+r_+)^2} \, \frac{1}{\om_+}
\int \limits_{0}^{+\infty} d\z \; \frac{ 1 + \a(\z)  }{ \cosh{\frac{\z}{\om_+}} -\cos{\frac{\pi}{\om_+}} } \; \frac{ e^{-\a(\z)} }{\cosh{\frac{\z}{2}}} 
\end{flalign}
where $ \a(\z) = 2 m (r'+r_+) \cosh{\frac{\z}{2}}$, and
\begin{flalign} \lb{f_wh}
f_{\xi} =
\frac{4q^2}{\pi w_+}
\sum\limits_{n=0}^{+\infty} \frac{1}{1+\delta_{n,0}}
\int \limits_{0}^{+\infty} dk \, K'_{\n}((r'+r_+) \sqrt{k^2+m^2}) W_{n}(k) \,. &
\end{flalign}

\begin{figure} [H] 
\centering
\captionsetup[subfigure]{margin={1.2cm,0.0cm},justification=centering}
\subfloat[Different couplings  ($\om=1 \mbox{ and } r_0 m = 1$)]
{\label{wh_mink}\includegraphics[width=0.42\textwidth]{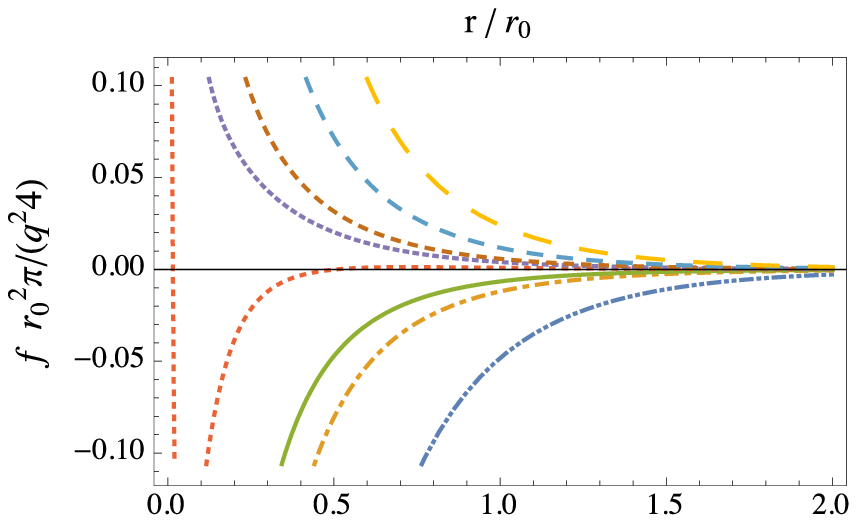}}
{\label{}\includegraphics[width=0.135\textwidth]{auxi}}
\subfloat[Different angle deficits ($r_0 m = 0.1 \mbox{ and } \xi= 0.1$)]
  {\label{sfwh}
    \includegraphics[width=0.42\textwidth]{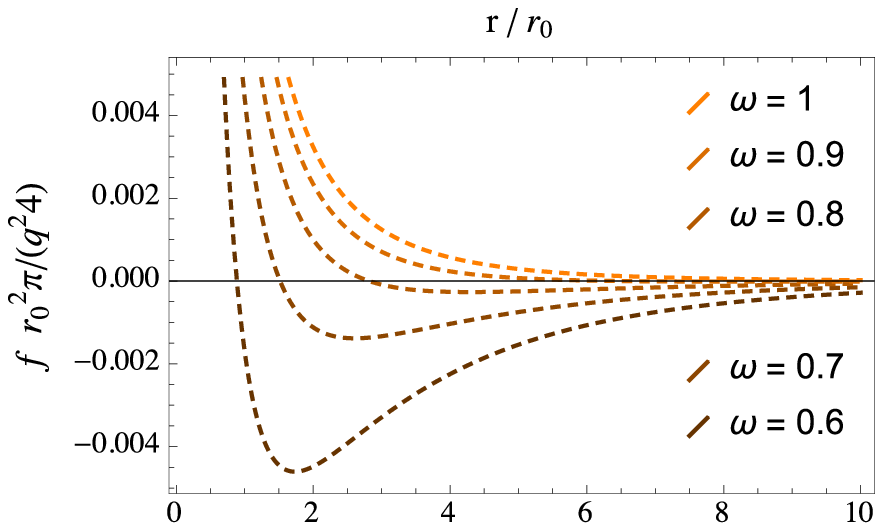}}
    \caption{{Dimensionless self-force $\frac{\pi r_0^2}{4q^2} f$ as a function of $r/r_0$ in cylindrically symmetric wormholes ($\om_{\mp} = \om$).}
    } 
      \label{sf_T2} 
    \end{figure}

\begin{figure} [H] 
\centering
\captionsetup[subfigure]{margin={1.2cm,0.0cm},justification=centering}
\subfloat[Minimal coupling]
{\label{ }\includegraphics[width=0.49\textwidth]{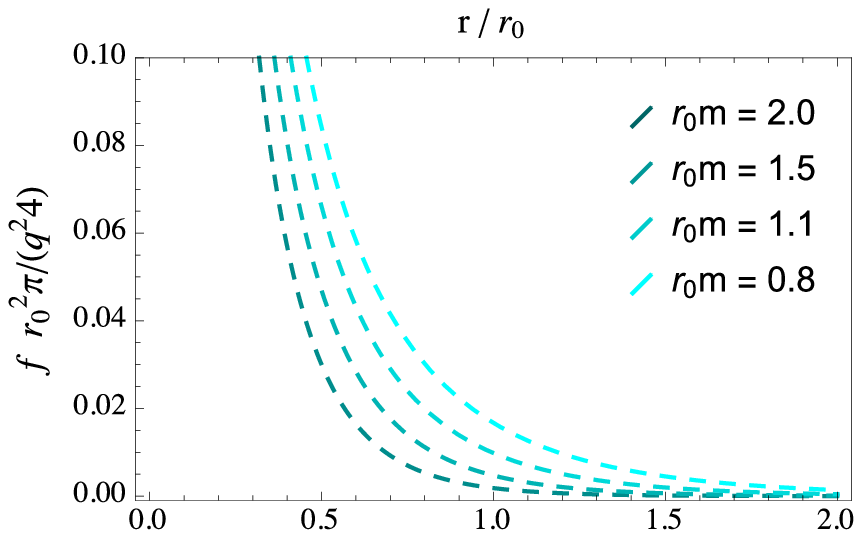}}
 \hspace{0.01cm}
\subfloat[$\xi = 0.3$]
  {\label{ }
    \includegraphics[width=0.49\textwidth]{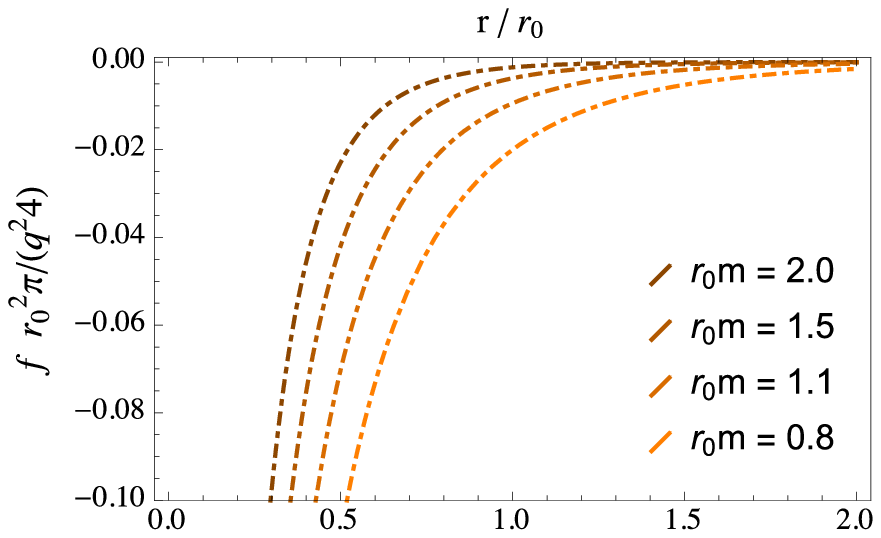}}
    \caption{{Dimensionless self-force $\frac{\pi r_0^2}{4q^2} f$ as a function of $r/r_0$ in a Minkowski cylindrical wormhole ($\om_{\mp}=1$) and different values of $r_0 m$.}
    } 
      \label{fmasawhmink} 
    \end{figure}

In order to complete the analysis of the results we consider symmetric wormholes across the throat, such that  $\om_- = \om_+ = \om$ and the throat radius is $r_0 = r_- = r_+ $.
Figures \ref{sf_T2} and \ref{fmasawhmink} show the dimensionless force $ \frac{\pi r_{i}^{2}}{4 q^{2}}  f$ plotted against the dimensionless position $ r/r_{0} $ of the particle for different spacetime configurations and coupling constants $\xi$ within the corresponding stability range (as studied in Figure \ref{f2}). 

Figure \ref{wh_mink} presents examples in a flat wormhole where $r_0 m = 1$. For minimal coupling the self-force is repulsive from the thin-shell throat as it was in the previous cases (Type I spacetimes) with positive $\k$. Negative values of the product $\xi \k$ produce an increased repulsion, while a positive product $\xi \k$ contributes with an attractive force towards the shell. At the specific value $\xi = 1/4$ the force changes sign and an attractive force at the vicinities of the throat is observed for greater couplings.
Figure \ref{sfwh} displays the self-force curves for different conical wormholes with $r_0 m =0.1$ and $\xi=0.1$ fixed.
In all the cases the force is repulsive in the vicinity of the throat and becomes asymptotically attractive if $\om<1$. 
Figure \ref{fmasawhmink} shows that, just as in Type I geometries, the self-force is attenuated as $r_0 m$ increases.


\section{Summary and conclusions}
\lb{s5}

The arbitrarily coupled massive scalar field of a charged particle held at rest in conical thin-shell backgrounds was solved to characterize stable and resonant solutions in terms of the configuration parameters.
We have determined a safety domain for the scalar Green's function and have applied it in the calculation of the self-interaction on a scalar charge, obtained from a regularized field at the position of the particle. 
The background geometries are locally flat except at the cylindrical shell\footnote{And at the central axis in case of a conical interior.}, so the field is coupled by $\xi$ to the  trace of the extrinsic curvature jump $\k$.
Two types of spacetimes were used to study the problem. 
Type I are conical geometries with the shell separating an interior from an exterior region characterized by negative or positive $\k$ (ordinary or exotic matter, respectively, at the shell).
Type II are cylindrical wormholes with a thin throat of positive $\k$ separating different asymptotic regions.

Resonances and poles were identified in the Green's function for the massive scalar field. 
Critical values $\xi_c^{(n)}$ of the coupling constant were found for which the $n^{th}$ mode of the Green's function is resonant. 
For given spacetime parameters and mass $m$ of the field, we determined a safety domain in terms of the coupling where no resonances or poles appear.
The latter is the stability domain of the solutions and restricts the allowed values of the coupling constant for scalar field models in the considered backgrounds, i.e., Type I geometries with flat or conical asymptotics, and Type II wormhole spacetimes.
As a general rule, $\xi > \xi_c^{(0)}$ is a stable range if $\k<0$, while $\xi < \xi_c^{(0)}$ are stable if $\k>0$. 
We found that the critical couplings for masses much greater than the reciprocal radius of the shell are $\xi_c^{(n)} \sim m/\k$.
Therefore, the stability regions can be enlarged, keeping the spacetime parameters fixed, by increasing the mass of the field.
On the other hand, the stable range decreases for smaller masses. If $m \to 0$ then $\xi_c^{(0)} \to 0$, and the safety domain takes only positive values of $\xi$ if $\k<0$, or only negative values if $\k>0$.
This last result coincides with the critical couplings found on a previous work were we used a massless scalar field
in conical backgrounds to test the self-force on a charged scalar particle \cite{cec17}. Despite this, we note that in the mentioned article some part of the analysis was performed outside the safety domain.

Stable couplings within the safety domain were used to study the self-force problem for a static scalar charge in Type I and Type II geometries. 
The self-force is the only force on the particle in these locally flat spacetimes.
Analyzing the problem in these simple backgrounds of the thin-shell type allows to isolate the effects of coupling to gravity at the shell and interprete how curvature affects the self-interaction directly from $\k$.
At minimal coupling the force is attractive to the shell in its vicinity if $\k<0$, it is repulsive if $\k>0$, and diverges as the particle gets closer to it.
As a general feature for non-minimal coupling we have observed that a negative product $\xi \k$ contributes with a repulsive force from the shell's position, and a positive value of $\xi \k$ contributes with an attraction towards the shell.
The coupling $\xi=1/4$, known for canceling out the delta-like effective potential at the shell 
for the radial solutions of the field \cite{cec17}, is manifested in the present work, i.e., with a negative (positive) $\k$ the force is attractive (repulsive) as the particle approaches the shell if $\xi<1/4$, and changes to a repulsion (an attraction) for $\xi>1/4$.
In all cases the absolute value of the self-force decreases with increasing mass as it was expected from the exponential decay $e^{-mr}/r$ in the general solution of the field in conical geometries. 
The same behavior was obtained in \cite{bezerra} in the case of a massive scalar field coupled to a spherically symmetric wormhole spacetime with infinitely thin throat.
In our cases, we observe that the field and force are localized at both sides of the shell either for Type I or Type II spacetimes.

\section*{Acknowledgments}
\lb{s6}
This work was supported by the National Scientific and Technical Research Council of Argentina (CONICET).

\end{document}